\documentclass[letterpaper]{llncs}

\usepackage{latexsym,amssymb,url}

\usepackage{times}
\usepackage{url}

\usepackage[linesnumbered,boxed]{algorithm2e}
\usepackage{amssymb}
\usepackage{amssymb}
\usepackage{amstext}
\usepackage{amsmath}

\usepackage{alltt}

\usepackage{graphicx}
\usepackage{rotate}
\usepackage{subfigure}

\newcommand{\xcomment}[1]{} % SI USA COSI': \xcomment{bla bla bla}

\newif\ifremarks
\remarkstrue %makes REMARKs visible
\begin{document}

\title{Design of QoS-aware Provisioning Systems}

%\title{SLA-driven Service Provisioning Systems}

\author{Michele Mazzucco\inst{1, 2} \and Manuel
Mazzara\inst{2} \and  Nicola Dragoni\inst{3}}
\institute{
Department of Computer Science, University of Cyprus
%\ \\email{michele@cs.ucy.ac.cy}
\and
School of Computing Science, Newcastle University, UK  %\\ \email{manuel.mazzara@newcastle.ac.uk}
\and
DTU Informatics, Technical University of Denmark %\\ \email{ndra@imm.dtu.dk}
}

\maketitle \pagestyle{plain}

% keywords: Utility Functions, Data Centers, Autonomic Systems, Web Services, Middleware

\begin{abstract}
We present an architecture of a hosting system consisting of a set of hosted
Web Services subject to QoS constraints, and a certain number of servers used
to run users demand. The traffic is session-based, while provider and users 
agree on SLAs specifying the expected level of service performance such that 
the service provider is liable to compensate his/her customers if the level 
of performance is not satisfactory. The system is driven by a utility function
which tries to optimize the average earned revenue per unit time.  The
middleware  collects demand and performance statistics, and estimates traffic
parameters in order to make dynamic decisions concerning server allocation and
admission control. We empirically evaluate the effects of admission policies, 
resource allocation and service differentiation schemes on the achieved 
revenues, and we find that our system is robust enough to successfully deal
with session-based traffic under different  conditions.
\end{abstract}

\section{Introduction}
\label{sec:introduction}

The increasing use of the Internet as a provider of services and a major 
information media has changed significantly, in the last decade, users 
expectations in terms of performance. It is simply considered no longer 
acceptable to wait a number of seconds to access a service or an information. 
Users, these days, expect a browser to perform like a home TV or a radio,
completely ignoring the basic differences between the functioning behind,  and
sometime cultivating unrealistic desires, especially regarding  failures
and robustness. This is one of the consequences of interpreting IT  systems as
information providers (due to the explosions of sites like Wikipedia  or
on-line encyclopedias) instead of a means for running calculations, as occurred
10 or 15 years ago.
A perfect example of this situation is described in~\cite{linder:2006a}:
%and~\cite{shankland:2008}: according to Google, an extra 0.5
according to Google, an extra 0.5 seconds in search page generation would kill user satisfaction, with a 
consequent 20\% traffic drop.
% while trimming the page  size
%of Google Maps by 30\% resulted in a traffic increase of 30\%.

%situation is the research done by Google: they found out that an extra 0.5
%seconds in search page generation would kill user satisfaction, with a 
%consequent 20\% traffic drop~\cite{linder:2006a}, while trimming the page  size
%of Google Maps by 30\% resulted in a traffic increase of
%30\%~\cite{shankland:2008}.

%\subsection{Performance-related Social Aspects and IT Consequences}

During the events of September 11, 2001 almost every news website became 
unavailable for hours \cite{lefebvre:2001} showing the weaknesses (but it
would be better to say the differences) of the Internet compared to the 
traditional information media. Situations like these certainly require a 
deeper investigation of the social impact of such approach in information 
retrieval (and ``keyboard dependency'') but this topic is definitely out of 
scope for this paper. Thus, given this ``embedded human behavior'', IT 
scientists can only interpret this need as a new challenge for performance 
requirements: customers expect not only resilience ({\it i.e.}, the capacity of
the system to recover from damage), but also performance. 
Besides, under-performing systems are rarely profitable.

These aspects easily trigger an important discussion regarding resilience and 
performance in this context. Given the higher and higher expectations in terms 
of performance, how will an average user be able to distinguish between a slow
service and a stuck or failed one? Although apart from~\cite{linder:2006a} we
are not aware of any other proper study in this field, it is not 
difficult to imagine that any under-performing system 
will simply be ignored and treated like a failed one. Thus, in the authors'
point of view, it would be simply unrealistic to consider resilience and
performance as two characteristics that can be analyzed separately. The nature of the problem 
forces us to consider Quality of Service (QoS) as part of system robustness.
Our opinion is that the issues related to service quality will eventually become
a significant factor in distinguishing the success or the failure of service 
providers. Being extremely difficult for service providers to meet the promised
performance guarantees in the face of unpredictable demand, one possible 
approach is the adoption of Service Level Agreements (SLAs), contracts 
specifying a level of performance that must be met and compensations in  case
of failure.

It is worth saying that the notions of compensations and failure here are 
different from the ones previously discussed by one of the authors  (for
example in \cite{MazzaraBPEL2009}). Here the compensation is a penalty to be
paid, while the failure is intended as a failure in meeting the specified
level of performance. In the previous works the compensation was instead a process  with
a designer-dependent logic with the goal of partially recovering a transaction 
made of a composition of different services. There are certainly analogies, but
a deeper investigation here is not possible due to space constraints. The
basic idea is that the theory presented in \cite{MazzaraL06} and the mechanism 
used there to dynamically trigger a compensation process can be exploited also 
to model the kind of scenarios presented in this work but with the evident 
open issue of time modelling.

\subsection*{Paper Contribution and Organization}

This paper addresses some of the performance problems arising when IT 
companies sell the service of running jobs subject to QoS, and thus
robustness, constraints. We focus on session-based traffic because, even 
though it is widely used ({\it e.g.}, Amazon or eBay), it is very difficult  to
handle; session-based traffic requires ad-hoc techniques, as job-based 
admission control policies drop requests at random and thus all clients 
connecting to the system would be likely to experience connection failures or
broken sessions under heavy load, even though there might be capacity on the 
system to serve all requests properly for a subset of clients. Also, since 
active sessions can be aborted at any time, there could be an inefficient use 
of resources because aborted sessions do not perform any useful work, but they
waste the available resources.

The contributions of the paper are threefold. First, we provide a formal model
describing the problem we want to tackle, that is to measure and optimize the
performance of a QoS-aware service provisioning system in terms of the average 
revenue received per unit time. According to this model, we then propose and 
implement an SLA-driven service provisioning system running jobs subject to 
QoS contracts. The middleware collects demand and performance statistics, and
estimates traffic parameters in order to make dynamic decisions concerning 
server allocation and admission control. The system architecture presented in
this work is based on Web Services technology and when we mention the word
``service'' we actually mean the specific technology. Anyway, this is just an
implementation choice that we will explain later. Other solutions would be 
certainly possible. Finally, we evaluate and validate our proposal through 
several experiments, showing the robustness of our approach under different 
traffic conditions.

The rest of the paper is organized as follows. Relevant related work is
discussed in Section~\ref{sec:related_work}, the problem we want to tackle is
formally modelled in Section~\ref{sec:problem_defintion} and policies for dynamic reconfiguration are discussed in Section~\ref{sec:policies}.
Section~\ref{sec:architecture} then presents the system's architecture and discusses
how the system deals with session-based traffic, while 
Section~\ref{sec:experiments} presents a number of experiments we have  carried
out. Finally, Section~\ref{sec:conclusions} concludes the paper highlighting
possible directions for future work.

\section{Related Work}
\label{sec:related_work}

There is an extensive literature on adaptive resource management techniques for
commercial data centers~({\it e.g.},~\cite{rajkumar:1997}, ~\cite{ghosh:2003},
~\cite{hansen:2004}). Yet, since previous work does not take into account the 
economic issues related to SLAs, service providers would still need to 
over-provision their data centers in order to address highly variable traffic 
conditions. Moreover, existing studies do not consider admission policies as a 
mechanism to protect data centers against overload
conditions~\cite{lefebvre:2001}.  However, as will become clear later in 
this paper, admission control algorithms have a significant effect on revenues.

The problem of autonomously configuring a computing cluster in order to satisfy
SLA requirements is addressed in several papers. Some of them consider the 
economic issues occurring when services are offered as part of a contract, 
however they do not address the problems affecting overloaded server systems 
({\it e.g.}, \cite{chandra:2003}, \cite{li:2005}, \cite{zhang:2004}), while
others include simple admission control schemes without taking any economic parameter into account. % ({\it i.e.}, \citep{menasce:2001, liu:2001}). For example, \cite{menasce:2001} propose an approach based on hill climbing techniques combined with analytic queuing models to guide the search for the best combination of configuration parameters of a multi-layered architecture hosting E-Commerce applications, while \cite{liu:2001} propose a theoretical model that uses both load balancing and server scheduling when trying to maximize the profit of a hosting platform subject to multi-class SLAs.

Finally, while there is an extensive literature on request-based admission
control ({\it e.g.},~\cite{urgaonkar:2005}, \cite{mazzucco:2007}),
session-based admission control is much less well studied. Also, nobody has 
studied the effects of combining admission control, resource allocation and 
economics when trying to model a commercial service provisioning system 
subject to QoS constraints. For example, \cite{villela:2007},
~\cite{urgaonkar:2005} and~\cite{levy:2003}  consider some economic models 
dealing with single jobs, but they focus on allocating server capacity only, 
while admission policies are not taken into account. Yet, revenues can be 
improved very significantly by imposing suitable conditions for accepting 
jobs. To our knowledge, the most closely related work is 
perhaps~\cite{mazzucco:2007}, that studies the effects of SLAs and  allocation
and admission policies on the maximum achievable revenues in the context of 
individual jobs. However, in E-business systems such as Amazon or eBay, 
requests coming from the same customer are related and thus they can be 
grouped into sessions. Unfortunately, if admission control policies like the 
one discussed in~\cite{mazzucco:2007} are in operation, a user trying to 
execute several related jobs would not know in advance whether all jobs will be
accepted or not. In this paper, instead, we implement some admission  policies
specifically designed to deal with session-based  traffic;
our approach uses a combination of admission control 
algorithms, service differentiation, resource allocation techniques and 
economic parameters to make the service provisioning system as profitable as possible.

\section{Problem Formulation}
\label{sec:problem_defintion}
In this section we present a mathematical model of the real world problem we 
intend to tackle. The reason for having a formal model is to abstract from the 
details we do not want to investigate, focusing only on those that are of 
interest for this work. The risk of formal models is always the over 
abstraction of problems; furthermore, interactions between aspects that are
included in the model and aspects that are excluded can complicate the 
situation. We intend to keep the model manageable and thus the proposed model is 
based on the concept of utility functions, a simple and common way for
achieving self-optimization in distributed computing systems. While different
kinds of utility functions can be employed, in this paper the average revenue obtained by  the
service provider per unit time is the considered metric. In a nutshell, the
model can be defined as follows: the user agrees to pay a specified amount for
each accepted session, and also to submit the jobs belonging to it at a 
specified rate. On the other hand, the provider promises to run all jobs
belonging to the session, and to pay a penalty if the average performance for
the session falls under a certain threshold.

More formally, the provider has a cluster of $N$ servers, used to run $m$
different type of services, while the traffic is session-based. A session is defined as follows:

\begin{definition}[Session]
 A session of type $i$ is a collection of $k_i$ jobs, submitted at a rate of $\gamma_i$ jobs per second.
\end{definition}

One strong assumption behind this model is the request of \textit{session
integrity} ({\it i.e.}, if a session is accepted, all jobs in it will be executed), as it is critical for commercial services. From a business perspective, the higher the number of completed sessions, the higher the revenue is likely to be, while the same does not apply to single jobs. Apart from the penalties resulting from the failure to meet the promised QoS standards, sessions that are broken or delayed at some critical stages, such as checkout, could mean loss of revenue for the service owners. From a customer's point of view, instead, breaking session integrity would generate a lot of frustration because the service would appear as not reliable. We assume that the QoS experienced by an accepted session of type $i$ is measured by the observed average waiting time:

\begin{equation}
 W_i = \dfrac{1}{k_{i}} \sum_{j=1}^{k_i} w_j \mbox{,}
\label{eq:avg_wt}
\end{equation}
where $w_j$ is the waiting time of the $j$th job of the session, {\it i.e}, the interval between its arrival and the start of its service.
%Each SLA agreement includes the following three clauses, namely charge, obligation and penalty.
Also, we assume that the provisioning contract includes an SLA specifying clauses related to charge, obligation and penalty.

\begin{definition}[Charge]
For each accepted session of type $i$, a user shall pay a charge of $c_i$.
\end{definition}

How to determine the amount of charge for each successfully executed session is outside the scope of this paper. However, intuitively, this could depend on the number of jobs in the session, $k_i$, and their submission rate, $\gamma_i$, or on the obligation. It is certainly expected that for stricter obligations there will be higher charges.

\begin{definition}[Obligation]
The observed average waiting time, $W_i$, of an accepted session of type $i$ shall not exceed $q_i$. 
\end{definition}

\begin{definition}[Penalty]
For each accepted session of type $i$ whose average waiting time exceeds the obligation ({\it i.e.}, $W_i>q_i$), the provider is liable to pay to the user a penalty of $r_i$.
\end{definition}

%Within the control of the service provider decides how many servers to allocate to each job type, and which session to accept and which to reject. The server allocation policy is invoked at session arrival and completion instants, while the admission policy is invoked every time a new session arrives. Of course, the allocation and admission policies are coupled: admission decisions depend on the amount of servers allocated to a certain job type and vice versa. Moreover, both of these policies should respond to dynamic changes in user demand. The problem is hot to do this in a sensible manner.

While the performance of computing systems can be measured using different metrics, in this paper we are interested in the average revenue received per unit time, as it is more meaningful from a business perpective than values such as the throughput or average response times.
 Thus, as far as the provider is concerned, the performance of the system is
 measured by the average revenue, $R$, received per unit time. That quantity
 can be computed using the following expression:

\begin{equation}
R = \sum_{i=1}^{m} a_{i} [c_{i} - r_{i}P(W_{i} > q_{i})] \mbox{.}
\label{eq:revenue_jobs_global}
\end{equation}

About Equation~\eqref{eq:revenue_jobs_global}, it is perhaps worth noting that
 while it resembles the utility function discussed in~\cite{mazzucco:2007}, here
 $a_i$ refers to the average number of type $i$ sessions that are accepted 
 into the system per unit time, while $P(W_i>q_i)$ is the probability that the
 observed average waiting time of a type $i$ session exceeds the obligation~$q_i$.
Also, while no assumption about the relative magnitudes of charges and
 penalties is made, the problem is interesting mainly if $c_i\leq r_i$.
 Otherwise one could guarantee a positive (but not optimal) revenue by 
 accepting all traffic, regardless of loads and obligations.
Finally, Equation~\eqref{eq:revenue_jobs_global} uses a ``flat penalty''
factor: if $W_i > q_i$ the provider must pay a penalty $r_i$, no matter  what
the amount of the delay is. Such a model can be easily extended.  For example,
one could introduce penalties that are proportional to the amount  by which the
waiting time exceeds the obligation $q$ (the effect of that would be to 
replace the term $P(W_i > q_i)$ in Equation~\eqref{eq:revenue_jobs_global} 
with $E(min(0, W_i - q_i))$).

%\begin{figure}
%\begin{center}
%\includegraphics[scale=0.4]{figures/utility_functions}
%\caption{Utility function with (a) flat penalties and (b) proportional
%penalties}
%\label{fig:utility_functions}
%\end{center}
%\end{figure}

Finally, instead of allocating whole servers to one of the $m$ offered services, the provider might want to share servers between different job types. If this is the case, it is possible to control the fraction of service capacity each service type is entitled to use, for example via block of threads. Those threads would thus play the role of servers.

\section{Policies for Dynamic Reconfiguration}
%\subsection{Policies}
\label{sec:policies}

Because of the random nature of Internet traffic and changes in demand pattern over time, accurate capacity planning is very difficult in the short time period and almost impossible in the long time period. On the other hand, if servers are statically assigned to the provided services, some of them might get overloaded, while others might be underutilized. It is clear that in such scenarios it could be advantageous to reallocate unused resources to oversubscribed services, even at the cost of switching overheads, either in terms of time or money.

The question that arises in that context is how to decide whether, and if so when, to perform such system dynamic reconfiguration. Posed in its full generality, this is a complex problem which does not always yield an exact and explicit solution. Thus, it might be better to use some heuristic policies which, even though not optimal, perform reasonably well and are easily implementable.
Within the control of the host are the ``resource allocation'' and ``job admission'' policies. 
The first decides how to partition the
total number of servers, $N$, among the $m$ service pools. That is, it assigns
$n_i$ servers to jobs of type $i$ ($n_1 + n_2 + \ldots + n_m = N$). The
allocation policy may deliberately make the decision to deny service to
one or more job types (this will certainly happen if the number of offered
services exceeds the number of servers).
The server allocation policy is invoked at session arrival and session completion
instants, while the admission policy is invoked at session arrival instants in
order to decide whether the incoming session should be accepted or rejected. 
Of course, the allocation and admission policies are coupled: admission
decisions depend on the allocated servers and vice versa. 
Moreover, they should be able to react to changes in user demand.

During the intervals between consecutive policy invocations, the number of 
active sessions remains constant. Those intervals, which will be referred to as
``observation windows'', are used by the controlling software in order to collect traffic 
statistics and obtain current estimates, as the queueing analysis carried out
at each configuration epoch requires estimates of the average arrival rates
($\lambda_{i}$) and service times ($b_{i}$), and squared coefficient of
variation of request interarrival ($ca_{i}^{2}$) and services times
($cs_{i}^{2}$).
Please note that all of the above parameters are time varying and stochastic in
nature, and thus their values should be estimated at each configuration
interval. However, if the estimates are accurate enough, the arrival rates and service times can be approximate as independent and
identically distributed (i.i.d.) random variables inside each window, thus allowing for online optimizations

 %of the average arrival rates and service 
%times, as well as the squared coefficient of variation, {\it i.e.}, the
%variance divided by the squared mean, of service and
%interarrival times. 
%(the advantage of using the squared
%coefficient of variation as a measure of variability, rather than the variance 
%or the standard deviation,  is that it is normalized by the mean, and so allows
%comparison of variability across distributions with different means). 
%These values are used by the policies at each decision epoch.

\begin{figure}[ht!]
\vspace{-5mm}
\centering
\includegraphics[width=0.7\textwidth]{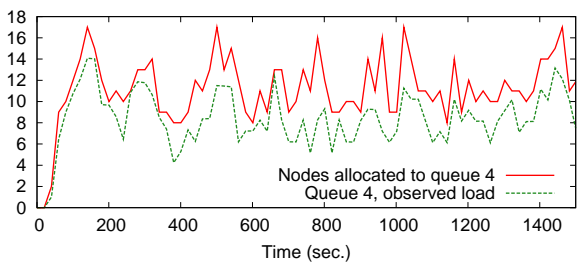} %architecture_new}
\caption{Dynamic resource allocation. Resources are allocated in proportion to the measured load.}
\label{fig:allocation}
\vspace{-5mm}
\end{figure}

In this paper, we implement and experiment with various heuristic policies. As
allocation algorithm we use the `Offered Loads' heuristic (see
Fig.~\ref{fig:allocation}), a simple adaptive policy that, using the traffic
estimates collected during the previous observation window, allocates the servers roughly in proportion to the offered
loads, $\rho_i=\lambda_i b_i$, and to a set of coefficients, $\alpha_i$,
reflecting the economic importance of the different job types (for service differentiation
purposes):

\begin{equation}\label{lalloc}
n_i = \left\lfloor N \frac{\rho_i \alpha_i}{\sum_{j=1}^{m}{\rho_j
\alpha_j}} + 0.5  \right \rfloor \mbox{,}
\end{equation}

\noindent (adding 0.5 and truncating is the round-off operation). Then, if the
sum of the resulting allocations is either less or greater
than $N$, adjust the number of allocated servers so that they add up to $N$.

For admission purposes we have embedded into our system three heuristics,
`Current State', 'Threshold' and `Long-Run'. The first two algorithms are
formally described in~\cite{mazzucco:2008}, and thus we only summarize them
here. The `Current State' policy estimates, at every arrival epoch, the changes
in expected revenue, and accepts the incoming session (possibly in conjunction
with a reallocation of servers from other queues) only if the change in expected
revenue is positive. In order to compute that value, it uses the state of each
queue, which is specified by the number of currently active sessions, the
number of completed jobs and average waiting time achieved so far (for each
session).

The `Threshold' heuristic uses a threshold, $M_i$, for each job type, and an
incoming session is accepted into the system only if less than $M_i$ sessions are active.
Each threshold $M_i$ is chosen so as to maximize $R_i$. We have carried out some
numerical experiments, and found that $R_i$ is a unimodal function of $M_i$. 
That is, it has a single maximum, which may be at $M_i = \infty$ for lightly
loaded systems. That observation implies that one can search for the optimal 
admission threshold by evaluating $R_i$ for consecutive values of $M_i$,
stopping either when $R_i$ starts decreasing or, if that does not happen, when the 
increase becomes smaller than some $\epsilon$. Such searches are typically very fast.

Finally, the `Long-Run' heuristic assumes that jobs of type $i$ are
submitted with the same arrival rate, that all sessions of type $i$ have the
same number of jobs, and that each queue is subject to a constant load of
sessions $L_i$. 
Suppose that queue $i$ is subjected to a constant load of $L_i$ streams
({\it i.e.}, as soon as one session completes, a new one replaces it) and has
$n_i$ servers allocated to it. Since each session consists of $k_i$ jobs
submitted at rate $\gamma_i$, the average period during which a session is
active is roughly $k_i/\gamma_i$ while, from Little's theorem, the rate at which
streams are initiated is $L_i \gamma_i/k_i$.
The above observations imply that, if over a long period, the numbers of active streams in
the system are given by the vector $L = (L_1, L_2, \ldots, L_m)$, and the server
allocation is given by the vector $n = (n_1, n_2, \ldots , n_m)$, the total
expected revenue earned per unit time can be computed using 
Equation~\eqref{eq:revenue_jobs_global}, where the average number of type $i$
sessions accepted per unit time, $a_i$, is replaced by $L_i \gamma_i / k_i$.

%If those assumptions are satisfied, then the average number of
%sessions accepted per unit time (see Equation~\eqref{eq:revenue_jobs_global})
%is simply $L_i \gamma_i / k_i$, and 

%Using observation windows to dynamically optimize the behavior of a computing
%system is not a novelty. However, while this paper proposes the use
%of event-based windows for optimization purposes, other scientists suggest
%time-based windows~\cite{bennani:2005} or complex algorithms  to detect traffic
%surges~\cite{bartolini:2008}. 
%Thus, the main difference is that our approach provides a simple 
%implementation of {\it adaptive windows}: under heavy traffic conditions  the
%optimization functions are executed more often then when the load is light and
%all requests would be accepted anyway.

\section{System Architecture}
\label{sec:architecture}

%Today's service provisioning systems are usually designed according to a
%three-tier software architecture. The first layer translates end-user markup
%languages into and out of business data structures, the second tier performs
%computation on business data structures while the third level provides storage
%functionality. Requests traverse tiers via synchronous communication over
%local area networks and a single request may revisit tiers more than once.
%Business-logic computation are often the bottleneck for Internet services, and
%thus this paper focuses on this tier. %However, user-perceived performance
%depends also on disk and network workloads at other tiers. Front-end servers
%are not typically subject to a very high workload, and thus over-provision is
%usually the cheapest solution to meet service quality requirements. Moreover,
%different solutions exist to address some of the issues occurring at both the
%presentation and database tiers~\cite{stewart:2005, doyle:2003}, 
%while~\citet{mcwherter:2004} have shown that smart scheduling can improve the
%performance of the database tier.

The three-tier software architecture presented in this work is based on Web Services technology. Web Services are self-describing, open components that support rapid, low-cost composition of distributed applications and their adoption looks like a promising solution to low cost and immediate integration with other applications and partners. The use of Web Services, in fact, eases the interoperability between different systems because they use open protocols and standards such as SOAP and HTTP. Computing systems are usually designed according to this three-tier software architecture (front-end, business logic and storage) but in this paper we focus mainly on the second one, as business logic computation is often the bottleneck for Internet services. Of course, user-perceived performance depends also on disk and network workloads at other tiers. However, front-end servers are not typically subject to a very high workload, and thus over-provision is usually the cheapest solution to meet service quality requirements, while different solutions exist to address some of the issues occurring at both the presentation and database tiers (see \cite{stewart:2005} and \cite{doyle:2003} for more details).
%while~\citet{mcwherter:2004} have shown that smart scheduling can improve the
%performance of the database tier.

\begin{figure}[!ht]
\vspace{-5mm}
\begin{center}
\fbox{
\includegraphics[width=0.7\textwidth]{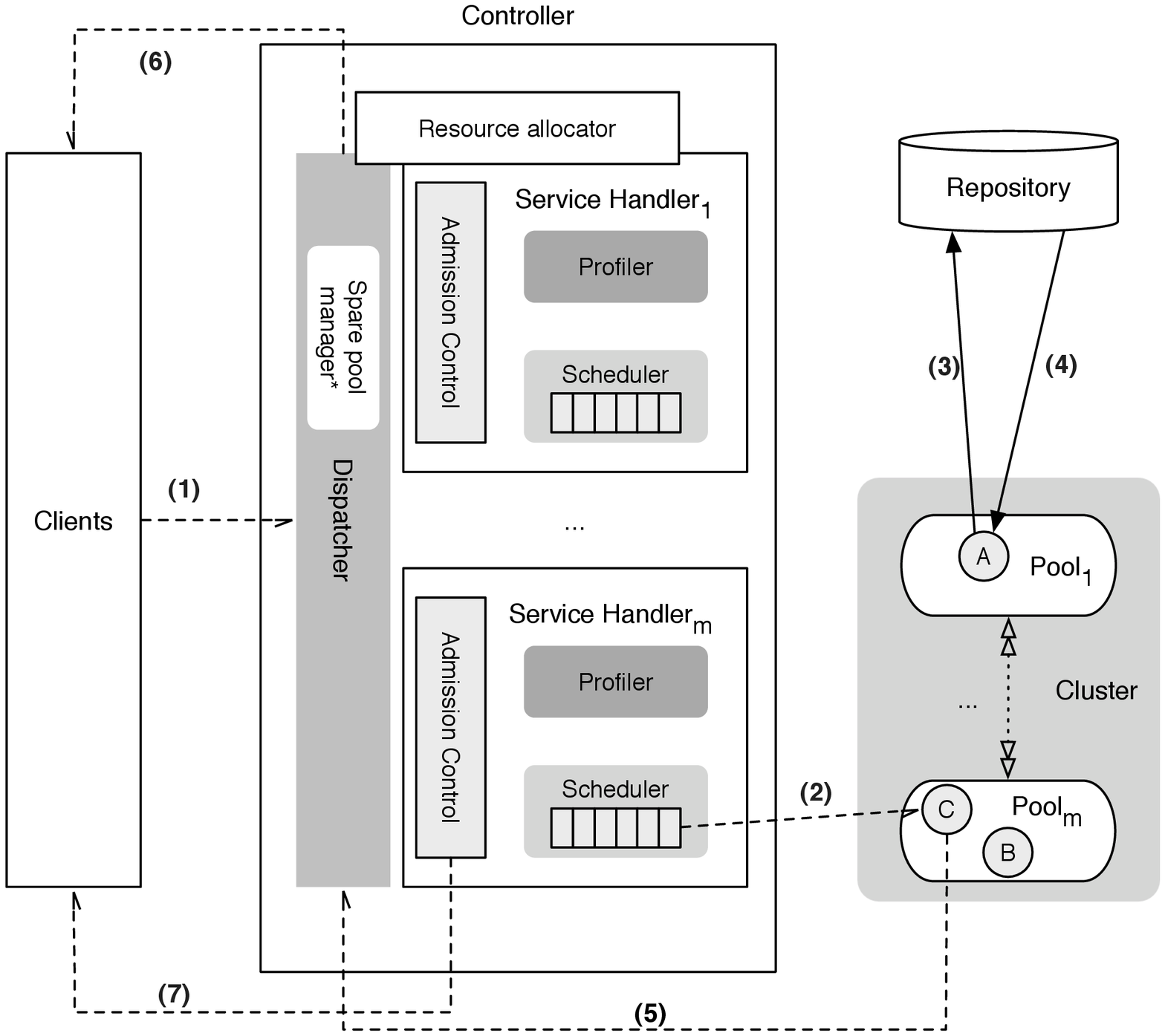}}
\caption{Architecture overview. Dotted lines indicate asynchronous messages.}
\label{fig:architecture_overview}
\end{center}
\vspace{-5mm}
\end{figure}

%\begin{figure}[ht!]
%%\vspace{-8mm}
%\centering
%\includegraphics[scale=0.35]{figures/architecture/arch_new}
%\caption{Architecture overview. Dotted lines indicate asynchronous messages.}
%\label{fig:architecture_overview}
%\vspace{-4mm}
%\end{figure}

The architecture we propose, shown in Figure~\ref{fig:architecture_overview}, uses a dedicated hosting model and follows the mediation service pattern~\cite{hohpe:2004}. The middleware  hides the IT infrastructure from the clients, creating an illusion of a single system by using a Layer-7 two-way architecture~\cite{cardellini:2002}, while the load balancer uses a
packet double-rewriting algorithm ({\it i.e.}, it forwards packets in both
directions, client-to-server and server-to-client) and takes rou\-ting
decisions using only the information available at the application layer of the OSI stack, such as target URL or cookie. This makes adding or removing servers at runtime straightforward, as clients do not know where their requests will be executed.
%The main advantage of Layer-7 over Layer-4 load balancers is that Layer-7 load balancers offer good policy management functionalities, {\it i.e.} the ability to define, integrate with existing policies and enforce, during the runtime cycle, policies such as access control and service level agreement compliance. Layer-4, instead, perform essentially a content-blind dispatching, which is faster and easy to implement, but less efficient because the employed routing algorithms are essentially stateless.
All incoming jobs are sent to the Controller (arrow 1), which performs the resource allocation, admission control and monitoring functions. For each type of service there is a corresponding Service Handler, which schedules incoming jobs for execution, collects traffic statistics through a profiler, and manages the currently allocated pool of servers. If the admission policy does not require global state information  ({\it e.g.}, threshold-based policies), then it too may be delegated to the Service Handlers.  If the same service is offered at different QoS levels and a threshold-based admission policy is employed, the Service Handler will be instantiated at dif\-fe\-ren\-tia\-ted service levels. Each level will have its own SLA management function instantiated that strives to meet that level of service specified by the differentiation. If the load is too high for any of the differentiated services, then the admission policy will start rejecting incoming traffic in order to maintain an adequate level of performance.  For policies that take into account the state of
all queues at every decision epoch ({\it i.e.}, state-based policies), instead,
there is no need to use different Service Handlers to deal with different QoS levels, as sessions can specify their own QoS requirements. %The advantage of this approach is that it  reduces the fragmentation during server allocations. Finally, the results of completed jobs are returned to the Controller, where statistics are collected, and then to the relevant user (arrows 5 and 6). If a deployment is needed, the service is fetched from a remote  repository (arrows 3 and 4). This incurs a migration time of a few seconds, while a server is switched between pools. Our system tries to avoid unnecessary deployments, while still allowing new services to be added at runtime. In such cases, the appropriate Service Handlers are automatically created in the Controller. If there is sufficient space on a server, deployed services could be left in place even when not currently offered on that server. One would eventually reach a situation where all services are deployed on all servers. Then, allocating a server to a particular service pool (handler) does not involve a new deployment; it just means that only jobs of that type would be sent to it. In those circumstances, switching a server from one pool to another does not incur any overhead. Our prototype implementation uses only one server for running the Controller infrastructure. Please note, however, that the controlling software can be easily clustered using techniques such as group communication or session replication in order not to become the system bottleneck or a single point of failure.

\subsection*{Session Arrival}

%\REMARK{MM}{Ho tolto l'UML, da sostituire con pseudocodice. La sezione salta
%ad un diverso livello di astrazione, eliminare i dettagli}

Here we discuss the steps taken at session arrival instants by state-based policies. 
In order to guarantee the correctness of the computation (multiple threads could see the
system in different states), consecutive requests are serialized by using a
pipeline with a single executor.
Every time a new session of type $i$ enters the system, the program sketched
in Algorithm 1 is executed. 
The algorithm first estimates the current
arrival rates and the potential arrival rates if the session was accepted
(the only arrival rate which changes is the one of queue $i$,
line~\ref{alg:new_arr_rate}), and simulates a new server allocation
using the potential arrival rates (line~\ref{alg:virtual_allocation}).
Then it computes the expected change in revenue, $\Delta R$.
The decision of accepting the new session, eventually with a reallocation of servers from other queues to
queue $i$, would increase the amount of charges by $c_{i}$, but it will also
increase the arrival rate at queue $i$ by $\gamma_{i}$. Thus, if the session was
accepted, there would be a possible penalty of $r_{i}$ in case the performance of
the new session was not met, and also different probabilities of paying
penalties for all the active sessions. 

\vspace{-5mm}

\begin{algorithm}[ht!]
 \SetAlgoNoLine
  \SetKwInOut{Input}{Input}
  \SetKwInOut{Output}{Output}
  \Input{A session arriving at queue $i$, $s_i$}
  \Output{The cookie for the session, if accepted, $-1$ otherwise}
  \BlankLine
 % \dontprintsemicolon
  \textbf{Phase I: Estimate $\Delta R$}\; 
  %(uses a pipeline with a single
  %executor to serialize the evaluation)}\;
  $(\lambda_{1}, \ldots, \lambda_{m}) \leftarrow
  \FuncSty{EstimateCurArrRate()}$\; \nllabel{alg:cur_arr_rate} 
  
  $(\lambda^{\prime}_{1}, \ldots,
  \lambda^{\prime}_{m}) \leftarrow (\lambda_{1}, \ldots, \lambda_{i-1},
  (\lambda_{i} + \gamma_{i}), \lambda_{i+1}, \ldots, \lambda_{m})$
  \; \nllabel{alg:new_arr_rate}
   
  $(n^{\prime}_{1}, \ldots, n^{\prime}_{m}) \leftarrow$
  \FuncSty{SimulateAllocation($\lambda^{\prime}_{1}, \ldots,
  \lambda^{\prime}_{m}$)}\; \nllabel{alg:virtual_allocation}

  $\Delta R \leftarrow c_{i} - [r_{i} \times 
  g(q_{i}, \lambda^{\prime}_{i}, k_{i}, n^{\prime}_{i})]$\;
  \nllabel{alg:delta_R_init}
  
  \For{$j \leftarrow 1$ \KwTo $m$}{ %\nllabel{alg:for_init}
 	\ForEach{session $t$ in queue $j$}{
 		$q_{j, t} \leftarrow \displaystyle{\frac{q_{j} k_{j} - u_{t} l_{t}}{k_{j} -
 		l_{t}}}$\;
 		
 		$\Delta g_{j} \leftarrow g(q_{j, t}, \lambda^{\prime}_{j}, k_{j}
 		- l_{t}, n^{\prime}_{j}) - 
 		g(q_{j, t}, \lambda_{j}, k_{j} - l_{t}, n_{j})$
 		\; \nllabel{alg:delta_g_j}
 		$\Delta R \leftarrow \Delta R - (r_{j} \times \Delta g)$\;
 	}
  } \nllabel{alg:delta_R_end}%\nllabel{alg:for_end}

%  $\Delta R \leftarrow$  \FuncSty{ac($(n_{1}, \ldots, n_{m}), 
 % (n^{\prime}_{1}, \ldots, n^{\prime}_{m}), 
  %(\lambda_{1}, \ldots, \lambda_{m}), 
  %(\lambda^{\prime}_{1}, \ldots, \lambda^{\prime}_{m}), 
  %session$)}\;
  
  \textbf{Phase II: generate the cookie and re-allocate servers}\;
  \eIf{$\Delta R > 0$}{ \nllabel{alg:accept_init}
  	$cookie \leftarrow$ \FuncSty{GenerateCookie($i$)}\;
  	\FuncSty{AddSession($queue_{i}, session$)}\;
  	\FuncSty{SetAllocation($n^{\prime}_{1}, \ldots, n^{\prime}_{m}$)}\;
  	\nllabel{alg:accept_end} }{
  	$cookie \leftarrow -1$\; \nllabel{alg:reject}
  }
  \Return{$cookie$}\;
\label{alg:session_arrival}
\caption{Session arrival, state-based policies.}
\end{algorithm}

\vspace{-5mm}

Denote by $g(x, \lambda, k, n)$ the
probability that the average waiting time for $k$ jobs exceeds the threshold
$x$, given that the arrival rate is $\lambda$ and that there are $n$ servers.
Having defined $g()$, the expected change in revenue resulting from a
decision to switch servers among queues and to accept a new session is computed in
lines~\ref{alg:delta_R_init}--\ref{alg:delta_R_end} as:

\begin{equation}
\Delta R = c_{i} - r_{i} g(q_{i}, \lambda_{i} + \gamma_{i}, k_{i},
n^{\prime}_{i}) - \sum_{j=1}^{m} r_{j} \sum_{t=1}^{L_{j}} \Delta
g_{j}(\cdot_{t})
\mbox{,}
\end{equation}

\noindent where $\Delta g_{j}(\cdot_{t})$ is the change in probability of
paying a penalty for session $t$ at queue $j$, see line~\ref{alg:delta_g_j},
while $L_{j}$ is the number of active sessions at queue $j$.
For session $t$ at queue $j$, the number of completed jobs is identified by
$l_{t}$, while the average waiting time over those jobs is $u_{t}$. Thus, the overall waiting
time that should not be exceeded over the remaining $k_{j} - l_{t}$ jobs,
trouble a penalty of $r_{j}$, is

\begin{equation}
q_{j, t} = \displaystyle{\frac{q_{j} k_{j} - u_{t} l_{t}}{k_{j} - l_{t}}}
\mbox{.}
\end{equation}

At the end of the for loop, if the expected change in revenue is positive the
new session is accepted, the cookie is generated, and server reallocation is put in operation
(lines~\ref{alg:accept_init}--\ref{alg:accept_end}). Otherwise, the session is
rejected and the server reallocation remains unchanged, see line~\ref{alg:reject}.

\section{Experiments}
\label{sec:experiments}

Several experiments were carried out in order to evaluate the robustness of our approach under different traffic conditions. 
However, because of space constraints, only some of them are discussed here. As
discussed in Section \ref{sec:problem_defintion}, the metric of interest is the
average revenue earned per unit time. CPU-bound jobs, whose lengths and arrival instants were randomly generated, queued and executed. We use synthetic load as this makes it easier to experiment with different traffic patterns. Moreover, we abstract from the hardware details such as number of cores or amount of memory; this way a job takes the same time everywhere, no matter on which hardware it is executed. Apart from the random network delays, messages are subject to random processing overhead, which cannot be controlled. Also, it could not be guaranteed that the servers were dedicated to these tasks, as there could be random demands from other users.
%The server capacity is guaranteed by a cluster of 20 (identical)servers running Linux with kernel 2.6.14, Sun JDK 1.5.0\_04, Apache Axis2 1.3 (to handle SOAP messages) and Tomcat 5.5 (to handle HTTP requests).
%The connection between the load generator and the controller is provided by a 100 Mb/sec Ethernet network, while the servers of the cluster are connected to the controller via a 1 GBit/sec Ethernet network. 
Each server can execute only one job at any time, {\it i.e.}, the system does not allow processor sharing (in Section~\ref{sec:conclusions} we suggest the possibility to extend the current system by running multiple jobs concurrently in a controller way in order to maintain the same QoS guarantees).
The scheduling policy is FIFO, with no preemption, while servers allocated to queue $i$ cannot be idle if there are jobs of type $i$ waiting. Finally, messages are sent using the HTTP protocol, as this is the most widely used protocol to exchange SOAP messages over the Internet.
In order to reduce the number of variables, the following parameters were kept fixed:
\begin{itemize}
\item The server capacity is guaranteed by a cluster of 20 servers offering four job types, {\it i.e.}, $N=20$, $m=4$.
\item The obligations undertaken by the provider are that the average observed waiting time of the session should not exceed the average required service time, {\it i.e.}, $q_i = b_i$.
\item All sessions consist of 50 jobs, {\it i.e.}, $k=50$. The job arrival rates are $\gamma_1 = \gamma_2 = \gamma_3 = 2$, while that for type 4 is $\gamma_4 = 1$. The average service time for all jobs is $b=1$.
\item Sessions are submitted with rate $\delta_1 = 0.1$, $\delta_2 = 0.04$ and $\delta_3 = 0.08$. 
\item The total offered load ranges between 60\% to over 100\% ({\it i.e.}, the
system would be overloaded if all sessions were accepted) by varying the
submission rate of type 4 jobs, $\delta_4 \in (0.02, 0.2)$.
\end{itemize}

%\subsection*{Markovian Traffic Scenario} 

In the following two experiments we assume that the traffic is Markovian, that
is, the sessions and jobs enter the system according to independent Poisson
processes, while service times are exponentially distributed.

\begin{figure}[!ht]
\vspace{-8mm}
\begin{center}
%\fbox{
\includegraphics{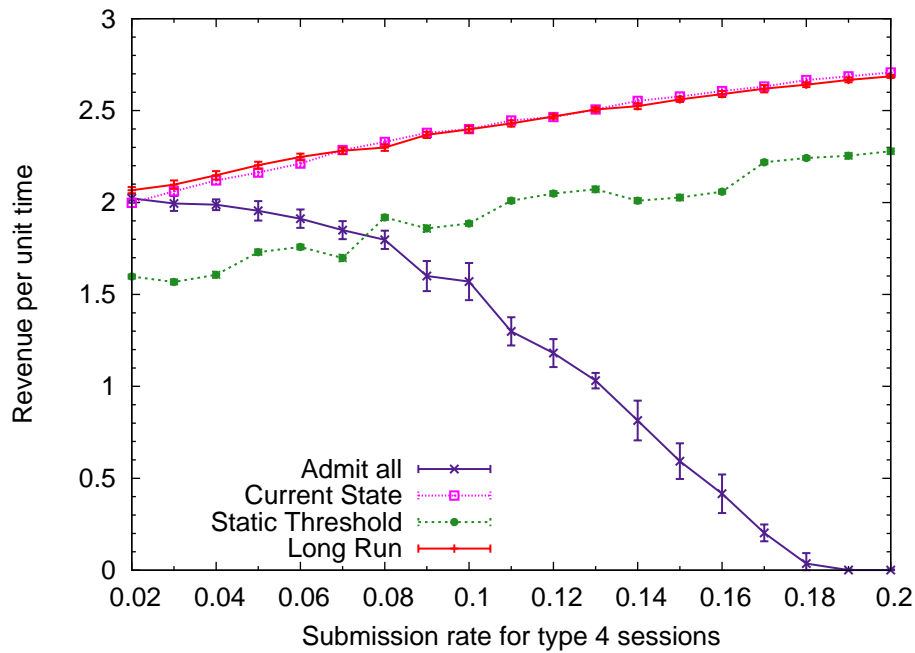}
%}
%\includegraphics[width=0.4\textwidth]{figures/ICSOC_experiments/fig4}}
\caption{[Experiment 1] Observed revenues when $c_i=r_i=10$, $\forall$~$i$.}
\label{fig:chap8_ 4_exp_c_eq_r}
\end{center}
\vspace{-8mm}
\end{figure}

%\subsubsection{Charges = Penalties.}
The first experiment, shown in Figure~\ref{fig:chap8_ 4_exp_c_eq_r}, measures
the average revenues obtained by the heuristic policies proposed in
Section~\ref{sec:policies} when all charges and penalties are the same, {\it
i.e.}, $c_i=r_i$, $\forall$ $i$: if the average waiting time exceeds the obligation, users get their money back. For comparison, the effect of not having an admission policy is also displayed.

Each point in the figure represents a run lasting about 2 hours. In that time, between 1,400 (low load) and 1,700 (high load) sessions of all types are accepted, corresponding to about 70,000 -- 85,000 jobs. Samples of achieved revenues are collected every 10 minutes and are used at the end of the run to compute the corresponding 95\% confidence interval (Student's $t$-distribution was used).
The most notable feature of this graph is that while the performance of the `Admit all' policy becomes steadily worse as soon as the load increases and drops to 0 when it approaches the saturation point, the heuristic algorithms produce revenues that grow with the offered load. According to the information we have logged during the experiments, they achieve that growth not only by accepting more sessions, but also by rejecting more sessions at higher loads.

%\paragraph{Other Metrics}

In some cases, values other than the average revenue per unit time might be of
interest. A possible example is the rate at which the sessions of type $i$ are
rejected, or the percentage of accepted sessions whose performance falls below
the minimum promised performance levels. For the the `Threshold'
heuristic, the former is given by:

\begin{equation}
X_{i} = \lambda_{i} p_{i, M_{i}} \mbox{,}
\end{equation}

\noindent where $p_{i, M_{i}}$ is the probability that a session of type $i$ is
rejected, that is, that there are $M_{i}$ active sessions in the $i$th queue
(no mathematical formula exists for the state-based policies).

\begin{figure}[ht!]
\centering
\subfigure[]{
\includegraphics[width=0.45\textwidth]{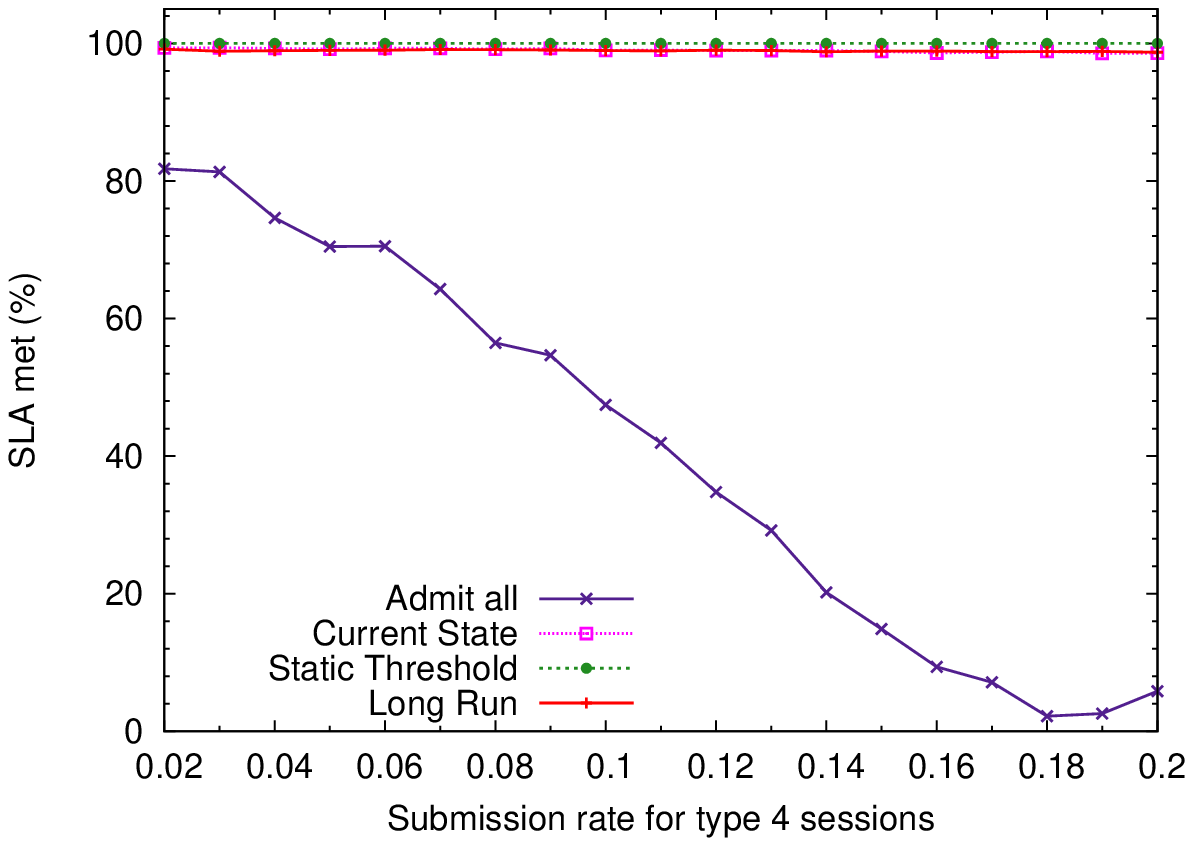}
\label{fig:qos}
}
\subfigure[]{
\includegraphics[width=0.45\textwidth]{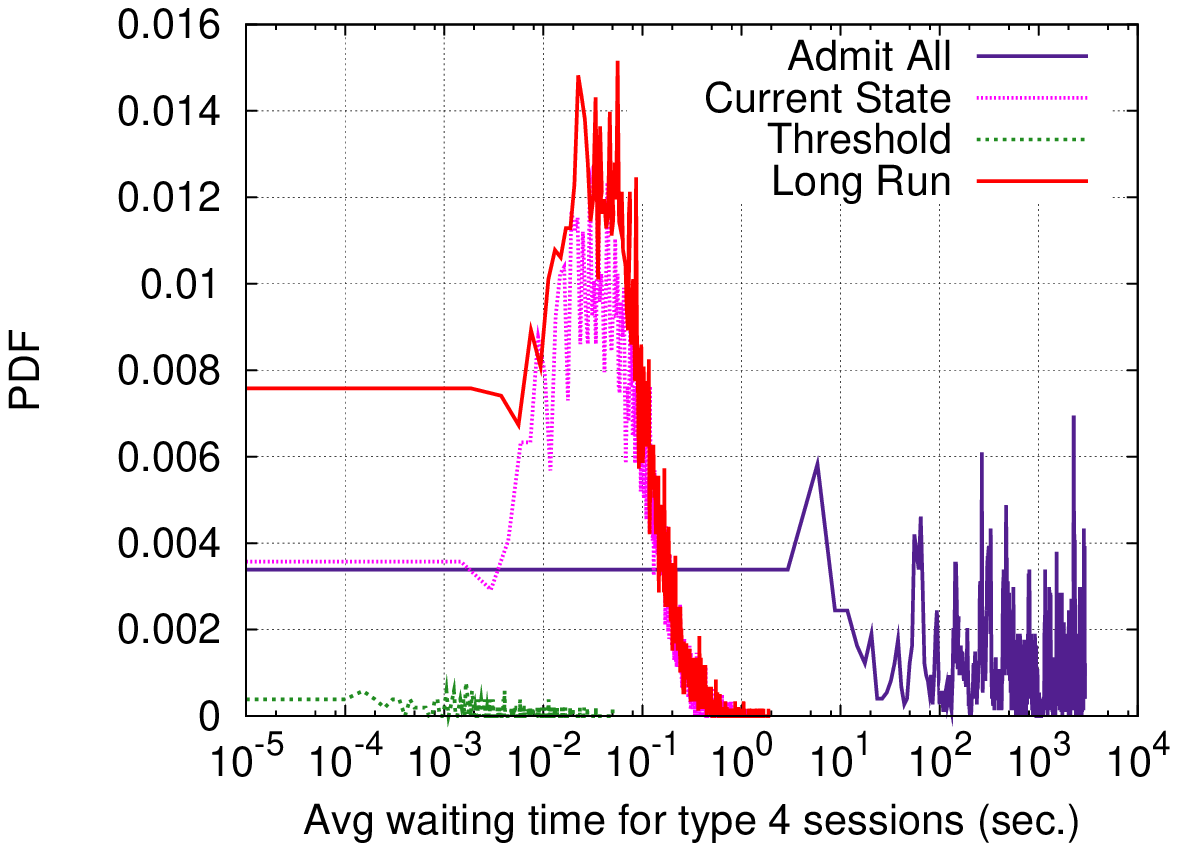}
\label{fig:pdf}
}
\caption{[Experiment 1] Other metrics: (a) SLA met for different policies, and
(b) Probability density function (PDF) of the observed average waiting
time for type 4 sessions, $\delta_{4} = 0.2$.}
\label{fig:other}
\end{figure}

Also, the performance of the various policies can be better understood by
observing other metrics; a policy might under-perform either because
it accepts too many sessions, thus failing to deliver the promised QoS
(like the `Admit All' policy in Fig.~\ref{fig:qos}), or because it rejects too
many sessions, thus missing income opportunities, like in the case of the
`Threshold' policy. Fig.~\ref{fig:pdf} shows very clearly that the `Threshold'
policy is very conservative, as almost all of the accepted sessions experience
an average waiting time of less than 0.1 seconds, while the minimum
acceptable performance level is set to 1 second.

\vspace{-5mm}

\vspace{-5mm}

\begin{figure}[!ht]
\vspace{-5mm}
\begin{center}
%\fbox{
\includegraphics[width=0.8\textwidth]{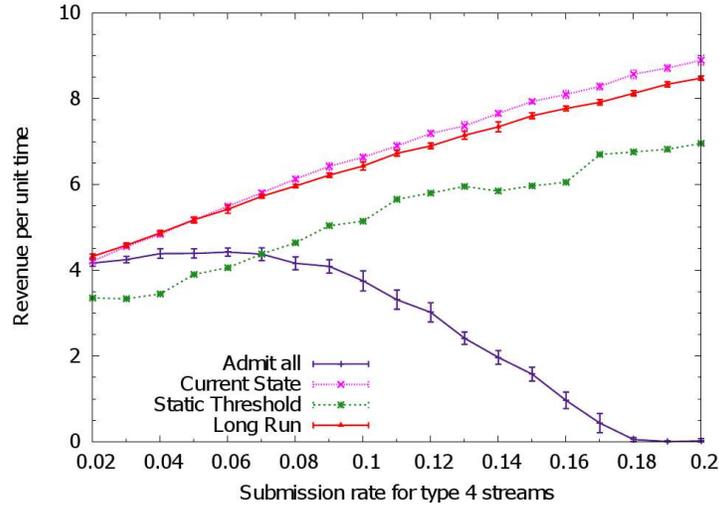}
%}
\caption{[Experiment 2] Observed revenues when $c_1=10$,$c_2=20$, $c_3=30$,$c_4=40$, $r_i=c_i$.}
\label{fig:chap8_ 4_exp_c_neq_r}
\end{center}
\vspace{-10mm}
\end{figure}

The second result concerns a similar experiment, except that now charges and 
related penalties differ between each job type:  $c_1= 10$, $c_2=20$, $c_3=30$ 
and $c_4=40$, $c_i=r_i$. The main difference compared to the previous 
experiment is that now it is more profitable to run, say, jobs of type 4  than
jobs of type 3. Figure~\ref{fig:chap8_ 4_exp_c_neq_r} shows that the maximum 
achievable revenues are now much higher than before in virtue of the higher 
charge values for type 2, 3 and 4 streams. Moreover, the `Long Run' heuristic
still performs very well, while the difference between the `Current State' and
the `Threshold' policies is about 25\%.

%\begin{figure}
%\vspace{-6mm}
%\begin{center}
%\includegraphics[scale=0.4]{figures/ICSOC_experiments/fig4}
%\caption{[Experiment 1] Observed revenues when $c_i=r_i=10$, $\forall$~$i$.}
%\label{fig:chap8_ 4_exp_c_eq_r}
%\end{center}
%\vspace{-1cm}
%\end{figure}

%\subsection*{High Variability Scenario} 

Next, we depart from the assumption that traffic is Markovian. A higher
variability is introduced by generating jobs with hyperexponentially 
distributed service times: 80\% of them are short, with mean service time  0.2
seconds, and 20\% are much longer, with mean service time 4.2 seconds.  The
overall average service time is still 1 second, but the squared
coefficient of variation of service times is now 6.15, {\it i.e.}. $cs^{2} =
6.15$. The aim of increasing variability is to make the system less predictable
and decision making more difficult. The charges are the same as in Figure~\ref{fig:chap8_ 4_exp_c_neq_r}, however if 
the SLA is not met, users get back twice what they paid, {\it i.e.}, $r_i =
2c_i$. The most notable feature of the graph shown in  Figure~\ref{fig:chap8_
4_hexp_2c_eq_r} is that now the revenues obtained by the `Admit all' policy 
become negative as soon as the load starts increasing because penalties are 
very punitive. On the other hand, the behavior of the three policies is 
similar. The `Current State' and `Long Run' algorithms performs worse than in 
the Markovian case (with $r_i = 2c_i$, not shown), while the wise `Thre\-shold' 
heuristic performs almost the same way.
Similar results were obtained in the case of bursty arrivals. They are not
shown here for the sake of space.

\begin{figure}[!ht]
\vspace{-8mm}
\begin{center}
%\fbox{
\includegraphics[width=0.8\textwidth]{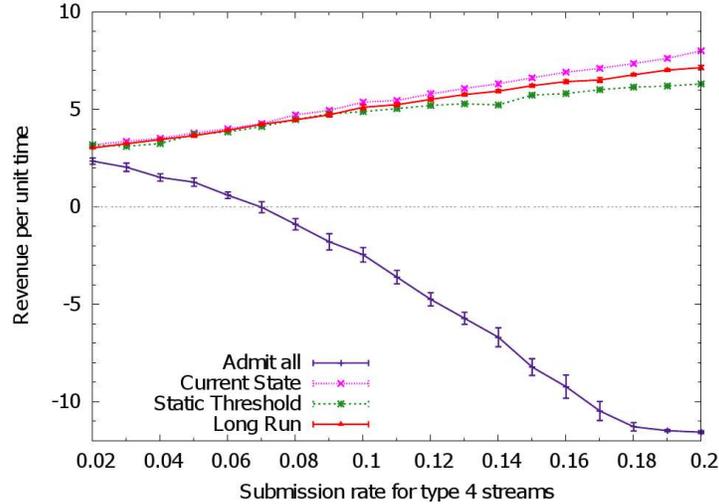}
%}
\caption{[Experiment 3] Observed revenues for different policies and two-phase
hyperexponentially distributed service times: $cs^{2}_{i}=6.15$, $r_i= 2c_i$,
charges as in Figure~\ref{fig:chap8_ 4_exp_c_neq_r}.}
\label{fig:chap8_ 4_hexp_2c_eq_r}
\end{center}
\vspace{-10mm}
\end{figure}

%\begin{figure}
%\vspace{-5mm}
%\begin{center}
%\includegraphics[scale=0.4]{figures/hexp/4_hexp_2c_eq_r}
%\caption{[Experiment 3] Observed revenues for different policies and hyperexponentially distributed service times: $cs^2=6.15$, $r_i= 2c_i$, charges as in Figure~\ref{fig:chap8_ 4_exp_c_neq_r}.}
%\label{fig:chap8_ 4_hexp_2c_eq_r}\end{center}
%%\vspace{-1cm}
%\end{figure}

%\vspace{-5mm}

\section{Conclusions and Future Work}
\label{sec:conclusions}

\vspace{-2mm}

In this paper we have presented a SLA-driven Service Provisioning System running jobs subject to QoS contracts. The system uses a utility function whose aim is to maximize the average revenue earned per unit time. We have demonstrated that policy decisions such as server allocations and admission control can have a significant effect on the revenue. %Moreover, those decisions are affected by the contractual obligations between clients and provider in relation to the QoS. 
The experiments we have discussed show that our system can successfully deal with session-based traffic under different traffic conditions. 
%The following are some directions for future research:
Possible directions for future research include sharing a server among several types of services or expensive system reconfigurations, either in terms of money or time (Amazon EC2, for example, can take up to 10 minutes to launch a new instance). Also in order to further improve the efficiency of the available servers, a concurrency level higher than one could be used. Of course, since the SLAs are still in operation, it is not possible to change the concurrency level at random: instead, the same QoS level as if jobs were ran alone should be maintained. Finally, one might want to increase the capacity of a data center by allowing it to be composed by several clusters. Such clusters may belong to the same organization or to different entities.

\small

\noindent \textit{Acknowledgments}: we would like to thank the EU FP7 DEPLOY Project (Industrial deployment of system engineering methods providing high dependability and productivity). More details at http://www.deploy-project.eu/.

%\begin{enumerate}
% \item Instead of having a dedicated architecture, it may be possible to share a server among several types of services.
%\item System reconfigurations, such as switching a server from one type of service to another, may incur non-negligible costs in either money or time (Amazon EC2, for example, can take up to 10 minutes to launch a new instance).
%\item In order to further improve the efficiency of the available servers, a concurrency level higher than one could be used. Of course, since the SLAs are still in operation, it is not possible to change the concurrency level at random: instead, the same QoS level as if jobs were ran alone should be maintained.
%\item One might want to increase the capacity of a data center by allowing it to be composed by several clusters. Such clusters may belong to the same organization or to different entities.
%\end{enumerate}

%\clearpage
%\begin{spacing}{0.5}
\bibliographystyle{abbrv}
\bibliography{eskdale}
%\end{spacing}

\end{document}